\documentclass[useAMS,usenatbib]{mn2e}
\usepackage{epsfig}

\title[H~1705--250]{Quiescent X-Ray/Optical Counterparts of the Black Hole Transient H~1705--250}
\author[Yi-Jung Yang et al.]{Y. J. Yang$^{1}$\thanks{E-mail:
y.j.yang@uva.nl (YJY)}, A. K. H. Kong$^{2}$, D. M. Russell$^{3}$, F. Lewis$^{4,5}$, R. Wijnands$^{1}$
\\
$^{1}$Astronomical Institute Anton Pannekoek, University of Amsterdam, Postbus 94249, 1090 GE Amsterdam, the Netherlands\\
$^{2}$Institute of Astronomy and Department of Physics, National Tsing Hua University, Hsinchu 30013, Taiwan\\
$^{3}$Instituto de Astrof\'isica de Canarias (IAC), v\'ia L\'actea s/n, La Laguna 38205, S/C de Tenerife, Spain\\
$^{4}$Faulkes Telescope Project, University of Glamorgan, Pontypridd, CF37 1DL, UK \\
$^{5}$Department of Physics and Astronomy, The Open University, Walton Hall, Milton Keynes, MK7 6AA, UK}
\begin{document}

\date{Accepted 2012 September 12. Received 2012 August 17}

\pagerange{\pageref{firstpage}--\pageref{lastpage}} \pubyear{2012}

\maketitle

\label{firstpage}

\begin{abstract}
We report the result of a new Chandra observation of the black hole X-ray transient H~1705--250 in quiescence. H~1705--250 was barely detected in the new $\sim$ 50 ks Chandra observation. With 5 detected counts, we estimate the source quiescent luminosity to be $L_X \sim 9.1\times10^{30}$ erg s$^{-1}$ in the 0.5-10 keV band (adopting a distance of 8.6 kpc). This value is in line with the quiescent luminosities found among other black hole X-ray binaries with similar orbital periods. By using images taken with the Faulkes Telescope North, we derive a refined position of H~1705--250. We also present the long-term light curve of the optical counterpart from 2006 to 2012, and show evidence for variability in quiescence.
\end{abstract}
\begin{figure*}
\psfig{file=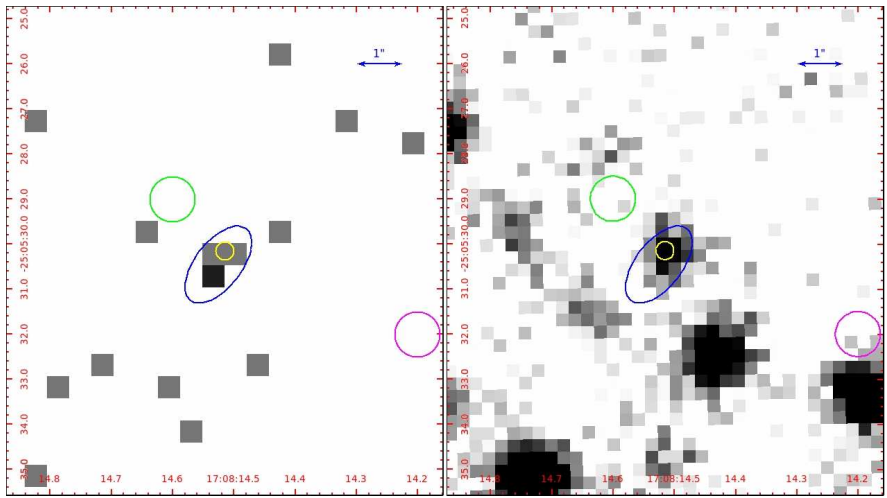,width=4.6in}
 \vspace{0.2cm}
 \caption{The left panel shows the Chandra image of the field around H~1705--250 in the 0.3-7 keV band. The right panel shows the Faulkes Telescope $i^{\prime}$-band image with the same field of view. The large blue ellipses in both images are the best-fit Chandra positions (with CIAO celldetect errors), and the small yellow circles are the best-fit optical positions derived from the Faulkes Telescope image compared with the 2MASS point source catalog (the errors are 0.2$\arcsec$, typical 2MASS point source precision). Green circles indicate the coordinates reported by Griffiths 1978, and the magenta circles indicate the coordinates from Remillard 1996 (the circles are 0.5$\arcsec$ in radius).}
\end{figure*}
\section{Introduction}
Soft X-ray transients (SXTs), or X-ray novae, are a sub-class of low mass X-ray binaries which contain a neutron star or a black hole primary accreting matter from the companion donor star via Roche lobe overflow. SXTs spend most of their lifetime in the quiescent state, and occasionally undergo dramatic outbursts which could last from weeks to months, and in some special cases, the outburst can go on for years (e.g. GRS 1915+105). The typical maximum outburst X-ray luminosity of such systems ranges from $10^{36}$ to $10^{39}$ erg s$^{-1}$, while the minimum luminosity can go as low as a few times of $10^{30}$ erg s$^{-1}$. However, the true quiescent luminosities (defined as the luminosities when no accretion onto the black hole occurs) are unclear due to the insufficient sensitivity of current instruments.

There are several theoretical attempts to explain the weak X-ray emission from X-ray binaries during their quiescent phase (see Narayan et al. 2001 for a review and references therein). The most widely used idea is perhaps through the advection dominated accretion flow (ADAF) model (Narayan \& Yi 1994, 1995, Narayan \& McClintock 2008 and references therein). The thermal energy that is created by the mass transfer is stored in the ADAF flowing towards the center of the compact object instead of being radiated away efficiently as it is at higher accretion rate (a thin disk is thought to be present in this case). Depending on the nature of the central compact objects, different quiescent luminosities are expected. In the case of a neutron star with a solid surface, the energy stored in the ADAF may impact onto the neutron star surface and be reradiated away. In the case of a black hole, no solid surface is present but an event horizon. A large fraction of the energy carried by the gas in the ADAF is then transported beyond the event horizon when the gas falls into the black hole before this energy could be emitted. The consequence of this scenario is that we expect to observe the quiescent luminosity of a black hole to be much fainter than a quiescent neutron star. With the sensitivity of current X-ray observatories (i.e. Chandra, XMM-Newton), we are able to observe this discrepancy and this provides strong evidence of the existence of the black hole event horizon (see Narayan \& McClintock 2008 for a review and references therein). In addition, there is also evidence showing that part of the energy could be dissipated as outflows moving away from the system resulting in low observed X-ray luminosities (Fender et al. 2003, Gallo et al. 2006).

H~1705--250 (also Nova Ophiuchi 1977, V2107 Oph) was discovered independently by both the HEAO-1 scanning modulation collimator and the Ariel 5 all sky monitor in September 1977 (Griffiths et al. 1978; Kaluziensku \& Holt 1977), and subsequently found to be associated with a bright 16.5 mag optical nova from observations taken at the Anglo-Australian Telescope and UK Schmidt Telescope(Longmore et al. 1997; Griffiths et al. 1978). A maximum X-ray flux (2-18 keV) of $\sim$3.5 Crab followed by a slow decline in the lightcurve was reported by Watson et al. (1978). Griffiths et al. (1978) also noted a dim object (at B$\sim$21) near the nova position on Palomar Sky Survey plates, which later was confirmed as the companion star (a K dwarf star with mass 0.3-0.6 $M_\odot$) of the binary system.

H~1705--250 has a confirmed dynamical measurement for the mass of its black hole in a range of 5.6-8.3 $M_\odot$ (Remillard \& McClintock, 2006), and it was not observed with any high sensitivity X-ray instrument prior our proposed Chandra observation. This is the first deep X-ray observation to measure the quiescent luminosity of this source. The last X-ray observation was taken with ROSAT and an upper limit of $5\times 10^{33}$ erg s$^{-1}$ (assuming a distance of 8.6 kpc) was placed (Narayan et al. 1997). Our 50 ks Chandra observation has improved the sensitivity by a factor of $\sim$1000.

In this paper, we present results of the X-ray observation and new optical monitoring data of H~1705--250 in quiescence. We also derive a more accurate source position which allows us to estimate the quiescent luminosity of the source more precisely.
\section[]{Data Analysis and Results}
H~1705--250 was observed with Chandra on 2010-05-02 UT 23:43:29 for a total duration of about 50 ks (PI: Kong, ObsID:11041). The observation was carried out with the Advanced CCD Imaging Spectrometer (ACIS) operating in the Very Faint mode, and the target was placed on a back-illuminated ACIS chip (S3). In additional to our X-ray observation, H~1705--250 has been monitored with the 2m Faulkes Telescope North since 2006. Most of the data were obtained using the SDSS $i^{\prime}$-band filter, and a few in Bessel V, R and Pan-STARRS y filters. The only solid detections of the SXT ($\gid 4\sigma$) were made in $i^{\prime}$-band.
\subsection{Astrometry Correction for both Faulkes images and Chandra position}
For the first examination of our Chandra data, we adopted the coordinates obtained from NED and SIMBAD (based on Griffiths 1978 optical nova position), which is inconsistent with the coordinates reported in Remillard et al. 1996. Within a 2$\arcsec$ extraction radius, there are 2 counts found at the Griffiths position, but no count at the Remillard position. However, we noticed a photon excess in between these two reported positions. We speculated that this might be the true position of H~1705--250, and a new astrometry correction for the position is necessary. We obtained a 200 s $i^{\prime}$-band image from the Faulkes Telescope North in 2010 (see Section 2.3), which was taken under excellent seeing conditions and the quiescent optical counterpart of the source is clearly detected. We tied the Faulkes image's world coordinate system (WCS) with the 2MASS point source catalog by using 'ccmap' task in IRAF. The task results in extremely small registration errors (0.064$\arcsec$ for RA, and 0.072$\arcsec$ for DEC). This provides us with good accuracy to determine the position of H~1705--250. The coordinates (J2000) we obtained are RA = 17:08:14.515, and Dec = -25:05:30.15. The typical precision of 2MASS point source astrometry is $<0.2\arcsec$ with respect to Tycho-2 reference system. We therefore use 0.2$\arcsec$ as our systematic position error.

To compare with the optical position, we also astrometrically corrected our Chandra image. We selected several nearby bright X-ray sources in which their positions were determined by using the task 'celldetect' in standard Chandra Interactive Analysis Observation (CIAO) package. We then compared the positions of these X-ray sources with their 2MASS counterparts, and updated the world coordinate system of our Chandra image. To check the consistency of the Chandra position of H~1705--250 with the optical position, we tried to detect the source using 'celldetect'. Because the source is very faint, we reset the default value of the parameter 'thresh' to 1 (the default setting is 3). We detected the source at position RA = 17:08:14.525, and Dec = -25:05:30.46 with errors 1.01$\arcsec$ and 0.51$\arcsec$, respectively. The Chandra position found by celldetect matches the optical counterpart position very well, and we can clearly see photons clustering at the overlapped region (see Fig. 1 for more detail).
\begin{figure*}
\rotatebox{-90}{\psfig{file=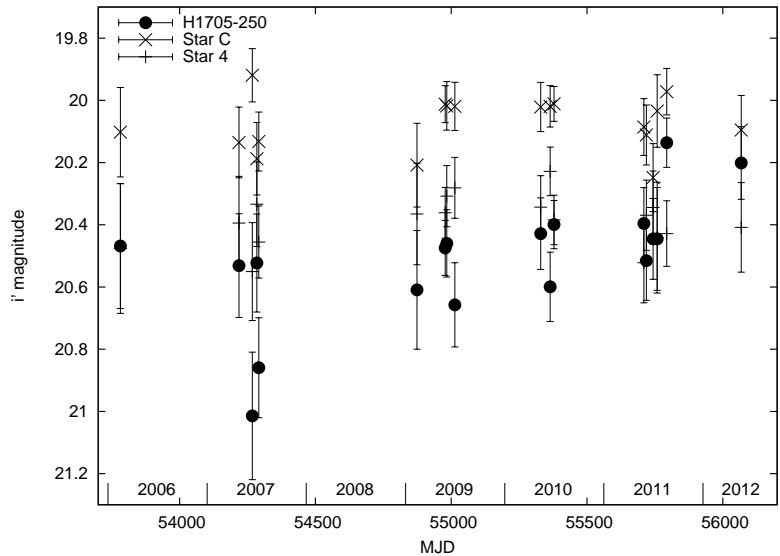, width=3in}}
 \vspace{0.2cm}
 \caption{Long-term lightcurve of H1705-250 spanning from 2006 to present. The optical lightcurve shows two dips in the epoch of 2007 and a brighter flux in the last two detections (2011 August -- 2012).}
 \label{sample-figure}
\end{figure*}
\subsection{Chandra}
We analyzed our Chandra data by using CIAO package version 4.3. In order to apply the most updated calibrations, we first ran the Chandra reprocessing script to create a new level 2 event file. The new level 2 event file was then used throughout the whole analysis. To reduce the background contamination and the calibration uncertainties, we extracted an energy filtered event list in the 0.3-7 keV energy band using the task 'dmcopy'. We created a source region file by selecting a circle with a radius 1.5$\arcsec$ centered at the new source position (derived from Faulkes data), as well as a background region file with much larger radius (10$\arcsec$) at a source free region near H~1705--250. We then ran the task 'dmextract' to obtain background subtracted source count rate. An alternative method utilizing built-in library funtool in DS9 was also performed to confirm the results by CIAO 'dmextract'. There were 5 photons found within the 1.5$\arcsec$ source region, and the average background within 10$\arcsec$ is around 0.304 counts. With 5 detected events, the Poisson distribution gives 95$\%$ confidence intervals from 1.6 to 11.7 counts. We calculated the Poisson probability of getting 5 photons at a random position to be $\sim8.5\times10^{-7}$, making the 5 photons detection significant. Because only a small number of photons were detected, we were not able to construct a spectrum for spectral analysis. The WebPIMMS was used to estimate the source flux by inputting the count rate obtained from CIAO, the Galactic column density $N_H$=$2.23\times 10^{21}$ cm$^{-2}$ along the line of sight towards the source position (Dicky \& Lockman, 1990), and assuming a power-law photon index $\Gamma = 2$ (a typical value for most of the black hole binaries in quiescent state, e.g. Kong et al. 2002). We obtained a flux of $1.03\times 10^{-15}$ erg cm$^{-2}$ s$^{-1}$ in the 0.5-10 keV energy band. Assuming a distance of $8.6 \pm 2$ kpc (Barret et al. 1996, Jonker \& Nelemans, 2004), we estimated the minimum luminosity of H~1705--250 to be $9.1\times 10^{30}$ erg s$^{-1}$. Since the distance uncertainty contribute the major errors, we estimated the lower and upper bound of the luminosity to be $5.4\times 10^{30}$ erg s$^{-1}$ and $1.4\times 10^{31}$ erg s$^{-1}$.
\subsection{Faulkes Telescope North}
H~1705--250 has been monitored since 2006 with the 2-m Faulkes Telescope North, located at Haleakala on Maui, as part of an ongoing monitoring campaign of $\sim 30$ low-mass X-ray binaries \citep{lewiet08}\footnote{http://faulkes-telescope/xrb}. 122 images of the source were taken between February 2006 and May 2012, on a total of 78 dates. Exposure times were 200 s and the pixel scale was 0.278 arcsec pixel$^{-1}$. Bias subtraction and flat-fielding were performed via automatic pipelines. Photometry was performed on H~1705--250 and five field stars using \small PHOT \normalsize in \small IRAF\normalsize.

The source was detected on 31 out of 78 dates with signal to noise ratio (S/N) $4 \leq S/N \leq 16$, all in the $i^{\prime}$-band filter. Images in which S/N $<5.5$ were discarded because the field star magnitudes were varying by $\geq 0.2$ mag, and images in which the seeing was $\geq 1.4$ arcsec were also removed due to the contamination of nearby stars to the aperture in this crowded field in the Galactic plane. The relative magnitudes were flux calibrated using Landolt standard stars observed in $i^{\prime}$-band on three dates in which the seeing was $< 1.7$ arcsec and the airmass of H~1705--250 was $< 1.5$. Five stars in the standard star field SA 110 were used on 2010-04-10 and 2011-07-16, and four stars in the field Mark A were used on 2010-07-09. All standards were taken within 4 hr of H~1705--250 and at a similar airmass ($< 0.3$ difference) on each date. The SDSS $i^{\prime}$-band magnitudes of the standard stars were calculated from their known $R$- and $I$-band magnitudes using the transformations of \cite*{jordet06}. Taking into account airmass-dependent atmospheric extinction, we calculated the magnitudes of three isolated stars of magnitudes $i^{\prime} = 16.3$ to 20.0 within 20 arcsec of H~1705--250 in order to calibrate our images. On the three dates, the derived magnitude of each star agreed within $\pm 0.02$ to 0.06 mag (the agreement was better for the brighter stars), confirming that the conditions were photometric on all three nights.

We then used the brighter two of these three stars\footnote{The faintest of the three stars was discarded because it was not detected with sufficiently high S/N in all images.} to perform relative photometry on H~1705--250 and two further faint field stars of similar magnitude to the SXT, which we designate star C and star 4. Star C is shown in fig. 1 of \cite{martet95}. We estimate the systematic error on the resulting magnitudes to be small; 0.03 mag. The magnitude errors are dominated by the low S/N of the SXT and faint field stars. In Fig. 2 we present the long-term Faulkes light curve of H~1705--250 and these two field stars of similar magnitude, spanning from 2006 to present.

The mean magnitude of H~1705--250, when detected significantly, was $i^{\prime} = 20.51$.
Mean quiescent apparent magnitudes in other filters measured at earlier epochs after outburst were $V = 21.5$ and $R = 20.8$ (Martin et al. 1995; Remillard et al. 1996).

The long-term optical quiescent variability of the SXT appears to be of higher amplitude than the two field stars (Fig. 2), which may be indicative of the modulation of the companion star or of accretion activity in quiescence, as seen in some other quiescent SXTs from optical or X-ray light curves \citep[e.g.][]{cantet10,cacket11,reymi11}. Flares on timescales of seconds to hours are often seen at optical wavelengths, indicating the presence of accretion activity even at these lowest luminosities \citep[e.g.][see also Section 3]{zuriet04,shahet05,shahet10}. The variability properties of the SXT and the two stars are given in Table 1. On two dates during 2007, H~1705--250 is visibly fainter than the mean by $\sim 0.4$ mag (dips), and it is brighter in the last two images with detections. The two dips and the brighter epoch at the end are probably real as they are $\geq 4 \sigma$ away from the mean magnitude (the rms scatter in the magnitude of star 4 is used as $\sigma$, since we assume this star to be intrinsically unvarying).

The magnitude of H~1705--250 spans 0.9 mag, whereas the field stars span 0.3 mag. This implies the variability in H~1705--250 is intrinsic to the source. Remillard et al. (1996) found an orbital periodic modulation in their $V$-band quiescent light curve, of full amplitude $\Delta V = 0.9$ mag and the folded light curve showed brief dips below the mean \citep[see also][]{martet95}. It is likely then, that the two dips in our $i^{\prime}$-band light curve are due to the same periodic modulation. The recent flux increase is harder to explain by the modulation; this may represent low-level accretion activity, as seen in other sources.

We also folded the data on the known orbital period of $0.5213 \pm 0.0013$ d (Remillard et al. 1996), and no phase-dependent variability is seen. Over the 6.3 years of observations, using the above period, the source will have performed $4385 \pm 11$ orbits during this time. Since the uncertainty is 11 periods, we cannot fold the light curve on the period and obtain meaningful results.
\begin{table}
\begin{center}
\caption{Optical variability properties of H~1705--250 and two field stars of similar magnitude. All values are given in $i^{\prime}$-band magnitudes.}
\begin{tabular}{lllll}
\hline
Star & mean mag & mean mag error & scatter & full range \\
\hline
H~1705--250 & 20.51 & 0.14 & 0.20 & 0.88 \\
Star C & 20.07 & 0.09 & 0.09 & 0.33 \\
Star 4 & 20.39 & 0.13 & 0.08 & 0.32 \\
\hline
\end{tabular}
\end{center}
\end{table}
\section{Conclusion}
We have observed the black hole transient H~1705--250 during its quiescent state with Chandra. The 50 ks long exposure reveals 5 photons at the source position (within 1.5$\arcsec$ radius), yielding a source luminosity of $L_X \sim 9.1\times10^{30}$ erg s$^{-1}$ (in 0.5-10 keV) when assuming a distance of 8.6 kpc. This improves the quiescent sensitivity of the source by a factor of $\sim 1000$ compared to the previous reported ROSAT value. In the context of the ADAF models, the quiescent luminosities depend on the orbital period of the system. The very low luminosity of H1705-250 together with its orbital period ($\sim 0.5$ days) thus consistent with this framework. We have also examined the optical monitoring data of H~1705--250 obtained by the Faulkes Telescope North, refined the best position of the source and obtained a long-term lightcurve.

It has been observed that the quiescent X-ray luminosity of black hole binaries are usually very dim (e.g. Garcia et al. 2001; Kong et al. 2002). They are $\sim$100 times fainter than neutron stars with similar orbital periods. It is unclear why the X-ray luminosity is much dimmer for black hole systems compared with the neutron star systems apart from the reason of a difference between a solid surface and an event horizon. The emission may originate from a truncated accretion disc that is detached from the central advective flow, or perhaps from the jet or outflow launched during quiescence or perhaps it is a combination for both disc and jet components. Several theoretical studies have attempted to explain the origin of the weak X-ray emission in the quiescent state of black hole binaries, yet none of the models can fully satisfy the observed properties. The ADAF is by far the most used scenario (as described in the introduction section). However, it has recently been realized that the jet/outflow can also play an important role in carrying away some fraction of the accretion energy from black holes in quiescence (Gallo et al. 2006), possibly resulting in weaker emissions comparing to emissions from neutron star systems. The true picture might fall in between the two scenarios (ADAF + outflows). Due to this, some theoretical models have evolved into more complicated forms to incorporate both disk and jet (or wind) properties (Blandford and Begelman 1999, Yuan et al. 2005). It is still a work in progress since most of the models are only able to explain some sources but not all. In addition, how do we know what we observed is the true quiescent state of the black hole? Due to current sensitivities of X-ray instruments, we are not able to observe most of the very faint black hole systems in detail and resolve their spectral properties in quiescence. However, several quiescent optical studies have revealed evidence for strong optical activities in quiescent lightcurves of several black hole systems (Zurita et al. 2003, Casares et al. 2009, Shahbaz et al. 2010). Most of these systems show short-term variability or flares superimposed on the weak ellipsoidal modulation of companion star, and sometimes show long-term aperiodic variability or magnitude color changes implying optical state changes (Cantrell et al. 2008).

From our 6 years long-term optical lightcurve, we observed two dips and an increasing brightness of the source since the end of 2011, indicating that the source is still active even at this very low rate of accretion. The origin of this variability is not yet fully understood, but it is probably associated with the accretion disc. There are several possible explanations: it could be due to the X-ray reprocessing in the accretion disc (Kong et al. 2001, Hynes et al. 2002, 2004), magnetic reconnection events (Zurita et al. 2003), or the emission from the ADAF (Shahbaz et al. 2003, 2010). Unfortunately, our optical data are not sufficiently sensitive for such detailed analysis. Future observations are needed for further investigation. In addition, the next generation X-ray observatory with improved sensitivity will certainly bring us new insights and shed light on understanding accretion physics of binary systems in very low accretion regimes.
\section*{Acknowledgments}
YJY and RW acknowledge support from the European Community’s Seventh Framework Programme (FP7/2007-2013) under grant agreement number ITN 215212 Black Hole Universe. AKHK acknowledges support from the National Science Council of the Republic of China (Taiwan) through grants NSC100-2628-M-007-002-MY3 and NSC100-2923-M-007-001-MY3, and the Kenda Foundation Golden Jade Fellowship. DMR acknowledges support from the Marie Curie Intra European Fellowship within the 7th European Community Framework Programme under contract no. IEF 274805. FL acknowledges support from the Dill Faulkes Educational Trust. RW acknowledges support from a European Research Council (ERC) starting grant. We thank Phil Charles and Tom Maccarone for helpful discussions, and Michiel van der Klis for reading the earlier version of this manuscript. The Faulkes Telescopes are maintained and operated by Las Cumbres Observatory Global Telescope Network.

\bsp

\label{lastpage}

\end{document}